# Mechanics of a Plant in Fluid Flow

Frédérick P. Gosselin

Laboratory for Multiscale Mechanics, Department of Mechanical Engineering, Polytechnique Montréal, Montréal, QC, Canada

Department of Botany, University of British Columbia, Vancouver, BC, Canada

frederick.gosselin@polymtl.ca

## Highlight

Plants reduce their drag by deforming in the flow, sway with the waves and make use of flow-induced vibrations to increase their exchange with the fluid they live in.

## Abstract

Plants live in constantly moving fluid, whether air or water. In response to the loads associated with fluid motion, plants bend and twist, often with great amplitude. These large deformations are not found in traditional engineering application and thus necessitate new specialised scientific developments. Studying Fluid-Structure Interactions (FSI) in botany, forestry and agricultural science is crucial to the optimisation of biomass production for food, energy, and construction materials. FSI are also central in the study of the ecological adaptation of plants to their environment. This review paper surveys the mechanics of FSI on individual plants. We present a short refresher on fluids mechanics then dive in the statics and dynamics of plant-fluid interactions. For every phenomenon considered, we present the appropriate dimensionless numbers to characterise the problem, discuss the implications of these phenomena on biological processes, and propose future research avenues. We cover the concept of reconfiguration while considering poroelasticity, torsion, chirality, buoyancy, and skin friction. We also cover the dynamical phenomena of wave action, flutter, and vortex-induced vibrations.

# Introduction

Wind buffets trees. Waves pound seaweeds. Plants live in constantly moving fluid, whether air or water. In response to the loads associated with fluid motion, plants bend and twist, often with great amplitude. They go with the flow (Denny and Gaylord, 2002). This strategy arising from evolution is very different from that favored by engineers when designing structures. Plants are "designed" for strength whereas in our daily life, we are more accustomed to objects designed for stiffness (Vogel, 2003). Although they achieve the same light collection function, leaves can deform and flutter under a wind gust, but solar panels are designed to keep their shape even in storm conditions. This design philosophy based on flexibility gives rise to Fluid-Structure Interactions (FSI) phenomena with large amplitudes not found in traditional engineering applications. New specialised scientific developments are thus required. FSI have been studied for a long time in engineering, typically to mitigate flow-induced vibrations problems (Blevins, 1990; Paidoussis, 1998; Naudascher and Rockwell, 2012). Many similarities exist between these engineering problematics and plant biomechanics because they are linked by the same physics, but plants grow and adapt to their environment and are typically more flexible (de Langre, 2008, 2012).

Studying FSI in botany, forestry and agricultural science is crucial to the optimisation of biomass production for food, energy, and construction material. Wind damages to cereal crops and forests are a major limiting factor in agricultural and forestry production. Through crop husbandry and forest management, lodging and windthrow can be limited and simulations of wind damages at the plant scale or at the canopy scale can help in optimising these husbandry and management techniques (Berry *et al.*, 2004; Mitchell, 2013; Dupont *et al.*, 2015; Gardiner *et al.*, 2016). To allow accurate predictions by the simulations, accurate accounts of the aerodynamic loadings on plants are required. This is where fundamental studies on FSI can have an impact in allowing an accurate understanding and evaluation of the underlying interaction mechanisms between vegetation and flows.

Fundamental biological issues require knowledge of the mechanisms of interaction between fluid flows and plants, notably for questions of adaptation. Over the long term, hydrodynamic phenomena exert evolutionary pressures which drive specialisation. A good example of this are the different strategies employed by seaweed to cope with drag: the "streamliner" seaweeds which minimise their drag and the "tolerators" which grow stronger (Starko and Martone, 2016). Within the lifespan of a plant, morphological plasticity allows vegetation and algae to adapt their growth to their habitat by responding to the mechanical stimuli of wind and water currents (Vogel, 1996; Koehl *et al.*, 2008; Coutand, 2010). At the core of these problems are the mechanisms of interactions between flow and plants.

Not only do wind and water currents affect plants, but plants also strongly affect the flow surrounding them. The presence and movement of plants alter the flow transport processes of momentum, heat and mass. In concrete terms, this alteration of transport influences the uptake of carbon dioxide by a forest (Finnigan, 2000), the sedimentation of particles such as nutrients and pollutants in marshes (Nepf, 2012), or the wave energy absorption of mangroves (Alongi, 2008). Moreover, leaf or blade movements influence their heat and mass exchange (Vogel, 2009; Rominger and Nepf, 2014), grass oscillations affect their pollen capture (McCombe and Ackerman, 2018), and wind-induced vibrations favour pollen release (Timerman *et al.*, 2014). These transport processes and many others depend on plant-flow interactions.

The goal of this review paper is to survey the mechanics of FSI on individual plants. We present the phenomena which arise from the coupling between the solid mechanics of a plant and the fluid mechanics of the wind or the water flow surrounding it. This review is specific to isolated plants, leaving out the literature on plant canopies (Finnigan, 2000; de Langre, 2008; Nepf, 2012). When they form forests, crop canopies, or seagrass covers, plants interact with wind or water flow to give rise to collective motions. These phenomena fall outside the scope of the present review. In our survey we present the statics and dynamics of plant-fluid interactions, focusing on an isolated plant. For every phenomenon considered, we present appropriate dimensionless numbers and discuss its implications on biological processes. We first provide a refresher of the basic fluid mechanics required to read this review. We introduce the concept of reconfiguration and then look at different complexifying aspects of it: poroelasticity, torsion, chirality, buoyancy, and skin friction. Secondly, we focus on dynamics, presenting in order phenomena related to wave action, flutter, and vortex-induced vibrations.

# Concepts of fluid mechanics

This section is a short refresher on fluid mechanics with an emphasis on dimensional analysis and drag, concepts required in the rest of the paper. Readers familiar with fluid mechanics can skip ahead to the next section. Those interested in getting a deeper introduction to fluid mechanics with a direct perspective into its biological implications are referred to *Life in moving fluids* (Vogel, 1996) or the textbooks of Denny (1988, 1993). White (2016) covers the material of a typical undergraduate fluid mechanics course in engineering, whereas Blevins (1990), Paidoussis (1998), Paidoussis *et al.* (2010), and Naudascher and Rockwell (2012) are reference textbooks on fluid-structure interactions.

A fluid is, by definition, a substance unable to resist shear, it cannot hold its shape. Whereas many substances such as mud, sand, and biological fluids can behave as fluids, we limit our treatment to air and water. In the range of problems considered here, air and water behave as incompressible and Newtonian fluids. This means that for fixed temperature and pressure, the density $\rho$ and the dynamic viscosity $\mu$ of a fluid are constants. For example, $at$ 20°C and 1 atm, air has properties $\rho = 1.20 \text{ kg} \cdot \text{m}^{-3}$ and $\mu = 1.80 \text{ E} - 5 \text{ kg} \cdot \text{m}^{-1}\text{s}^{-1}$. Under similar conditions, fresh water has properties $\rho = 998 \text{ kg} \cdot \text{m}^{-3}$ and $\mu = 1.003 \text{ E} - 3 \text{ kg} \cdot \text{m}^{-1}\text{s}^{-1}$ (White, 2016).

Although air and water have densities and viscosities different by orders of magnitude, they can be observed to behave in the same way under the right conditions. This is the principle of similitude. Let's consider a body with a characteristic dimension $\ell$, subjected to a flow of velocity $U_\infty$ of a fluid of density $\rho$ and viscosity $\mu$. Through dimensional analysis, considering the units of the quantities at hand, we can reduce the number of parameters. The Buckingham theorem (White, 2016 Chap. 5) states how many dimensionless parameters are necessary to describe the problem. Here we have $n = 4$ parameters ($\ell, U_\infty, \rho, \mu$), and $k = 3$ fundamental dimensions, namely length [m], time [s] and force [N] (keeping in mind that mass is not independent since $1\text{N} = 1 \text{ kg} \cdot \text{m} \cdot \text{s}^{-2}$). The Buckingham theorem tells us that we can formulate $n - k = 1$ dimensionless number to fully characterise this problem. Many dimensionless numbers could be defined to simplify the quantities of this problem, but it's customary to use the well-known Reynolds number

$$Re = \frac{\rho \ell U_\infty}{\mu}. \quad (1)$$

The Reynolds number is an indication of the ratio of inertial ($\sim \rho \ell^2 U_\infty^2$) to viscous forces ($\sim \mu \ell U_\infty$) in the flow. At low values, viscous forces dominate and one consequence of this is that flows tend to be laminar, i.e. flow streamlines stay well organised and smooth. By opposition, at high Reynolds numbers, inertial forces dominate, and flows become turbulent with eddies and fluctuations.

By similitude, if the Reynolds number is matched between two experiments, the same flow with similar structure and form should be observed. For example, a ball of size $\ell = 0.05$ m, flying in air at $U_\infty = 30 \text{ m} \cdot s^{-1}$ has a Reynolds of $Re = 10^5$. To reach the same Reynolds number in water, that same ball would have to travel at $U_\infty = 2.01 \text{ m} \cdot \text{s}^{-1}$. By achieving similitude through matching Reynolds numbers, we can expect the flow structure in these two experiments to be the same, i.e., laminar or turbulent conditions, similar wake, similar boundary layer, etc.

In this ball example, we can achieve similitude with matching a single dimensionless number ($Re$), but more complicated problems with more parameters (like gravity, a free surface, structural flexibility, etc.) require matching more dimensionless numbers. We see many examples of this further in this review.

Drag is the force a fluid flow exerts on a body in the direction of the flow.[1] The drag of a static body subjected to a uniform fluid flow of constant velocity (or equivalently, a body moving at constant speed in a stationary fluid) is the sum of two components: skin friction drag, and pressure drag (Hoerner, 1965). Skin friction is proportional to viscosity and wetted surface, whereas pressure drag is proportional to the momentum carried by the flow and the cross-section of the body normal to the flow. The relative importance of these forces depends on the Reynolds number and on the type of body. For low Reynolds numbers and for streamlined bodies, skin friction drag is most important. For pollen transport, flying seeds or for long kelp blades aligned with the flow, skin friction is dominant. It acts as shear parallel to the surface of the body. On the other hand, at high Reynolds numbers and for bluff bodies (non-streamlined bodies which generate a significant wake), pressure drag is most important. As the name implies, pressure acting normal to the object's surface adds up to create the drag.

Returning to our body subjected to flow with parameters $(\ell, U_\infty, \rho, \mu)$, we now consider its drag $F_R$ combining skin friction and pressure drag as it would be measured with a dynamometer or a force balance in a wind tunnel or tow tank. From these $n = 5$ parameters and $k = 3$ fundamental dimensions, Buckingham tells us that we can describe the problem with $n - k = 2$ dimensionless numbers. In addition to the Reynolds number, a second number is required. We define the drag coefficient:

$$C_D = \frac{F_R}{\frac{1}{2}\rho a \ell^2 U_\infty^2}, \quad (2)$$

where $a$, is a constant geometrical parameter. The drag coefficient is a dimensionless way to represent drag. It depends on the shape of the body in question, down to its surface roughness, and it also depends on the Reynolds number. Hoerner (1965), White (2016) and many others present curves of $C_D(Re)$ for standard bodies. Eq. (2) can be manipulated to isolate the drag

$$F_R = \frac{1}{2}\rho A C_D U_\infty^2, \quad (3)$$

where we replaced the generic dimension $\ell$ by the cross-sectional area $A = a\ell^2$. To estimate the drag of a body of a standard shape, one first needs to estimate the Reynolds number, and then look up in a fluid mechanics textbook or handbook for the appropriate $C_D$ value. This works well for rigid engineering constructions, but for plants which tend to be highly flexible and which deform under flow, the dimensional analysis is more complicated as elasticity enters the problem.

## Reconfiguration

Plants rely on flexibility to change form and reduce their drag when subjected to fluid flow. Rather than saying that plants deform, which carries a pathological meaning, Vogel (1984) coined the term *reconfiguration* to describe the change of shape of plants under fluid flow. By bending and twisting, tree leaves, branches, and whole crowns, as well as macrophytes or seaweed blades reduce their cross-sectional area normal to the flow and become more streamlined (Vogel, 1989; Sand-Jensen, 2003; Harder *et al.*, 2004; Rudnicki *et al.*, 2004; Vollsinger *et al.*, 2005; Martone *et al.*, 2012). At high Reynolds number, a typical bluff body perceives a drag force proportional to the square of the flow velocity. Because plants reconfigure, their drag deviates from this scaling. We express this deviation with the Vogel exponent (Vogel, 1984, 1996):

$$F \propto U_\infty^{2+V}, \quad (4)$$

where $F$ is the drag force, $U_\infty$ the freestream flow velocity away from disturbances, and $V$ the Vogel exponent.

---

[1] The force it exerts perpendicular to the flow is the lift, but it is not essential in the present review.

For example, the tuliptree leaf, *Liriodendron tulipifera*, reconfigures by rolling into an increasingly acute cone under increasing wind speed (Vogel, 1989). This reduces its cross-section normal to the flow and makes it more streamlined. Whereas if the leaf were rigid, its drag would scale with the square of the flow velocity ($V = 0$), Vogel (1989) evaluated a more or less linear increase of drag with flow velocity ($V \sim -1$) from wind tunnel measurements. Similarly, the crown of a Murray pine, *Pinus contorta*, bends and folds back on itself under high winds (Rudnicki *et al.*, 2004). The tree crown reduces its cross-sectional area, it becomes more streamlined as its leaves align with the flow, and it becomes more compact. The latter effect leads to a reduction of porosity inside the tree crown, possibly slowing down the flow.

What comes across from these two examples of the leaf and the tree crown is that three mechanisms are responsible for drag reduction by reconfiguration: area reduction, streamlining, and effective velocity reduction. To mathematically represent these three mechanisms, we rewrite Eq. (3) to consider the drag on a *flexible* object

$$F = \tfrac{1}{2}\rho A_f C_{Df} U_f^2, \quad (5)$$

where $A_f$ is the cross-sectional area of the deformed object normal to the flow, $C_{Df}$ is the drag coefficient of the deformed object, and $U_f$ is the effective flow velocity perceived by the object. The subscript $f$ is there to remind us that the quantity depends on the change of shape of the object. The three mechanisms of drag reduction can be quantified by evaluating how the area reduces $A_f(U_\infty)$, the body streamlines $C_{Df}(U_\infty)$, and the effective velocity changes $U_f(U_\infty)$ as functions of the freestream flow velocity. On a simple shape like a fibre, a sheet, or the leaf of Figure 1, drag reduction occurs only through the mechanisms of area reduction and streamlining. Figure 1 illustrates these two mechanisms by comparing a low flow speed $U_{\infty a}$ case in (a) and a high speed $U_{\infty b}$ in (b). The leaf perceives the unmodified free stream flow velocity $U_f = U_{\infty a}$ or $U_f = U_{\infty b}$ in Figure 1 (a) and (b), respectively. Therefore, the mechanism of effective velocity reduction does not play a role here. The cross-sectional area of the leaf reduces as it bends down with the flow further. Moreover, its geometry becomes mores aligned with the flow and more streamlined. Its speed-specific drag coefficient reduces with wind speed, i.e., $C_{Dfb} < C_{Dfa}$. We first turn our attention to these two mechanisms and later return to discuss effective velocity reduction.

### Area reduction and streamlining

Theoretical interpretation is difficult to obtain from real plants. Their geometries are complex and change with time as the plant grows, their materials have anisotropic properties varying with water content, and no two specimens are ever identical. To better understand the reconfiguration mechanisms and develop new analytical tools, it is easier to consider idealised structures such as beams, rods, and plates made of engineering materials. A flat rectangular sheet (cut out from an overhead projector transparency film) held at its centre and placed perpendicular to the flow in a wind tunnel exhibits the same mechanisms of reconfiguration as a natural leaf (Gosselin *et al.*, 2010). The schematics of Figure 2 (a) shows how the plate bends down in the flow, reducing its cross-sectional area, and streamlining its shape. Figure 2 (b) shows the drag measurements of three plates of identical dimensions with decreasing rigidty: very large (blue squares), small rigidity (white circles) and smallest rigidity (green diamonds). Whereas the drag of the rigid specimen obeys a $U_\infty^2$ law, that of the flexible specimens increase in a less pronounced way and reach smaller magnitudes.

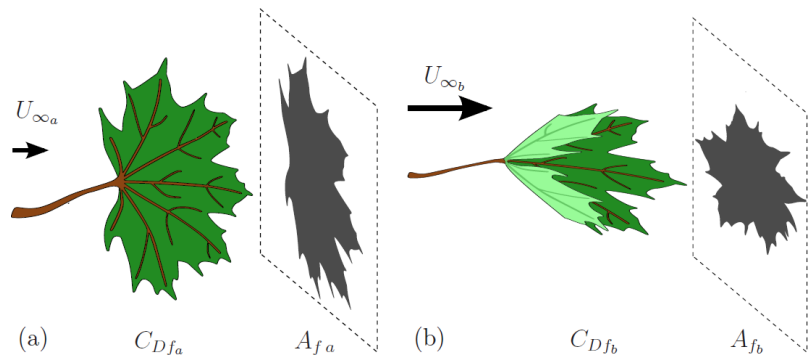

Figure 1 Schematics of the drag reduction on a generic leaf. At low speed, the leaf is little deformed (a) and at high speed it reconfigures (b), hence decreasing its cross-sectional area normal to the flow $A_{fb} < A_{fa}$, and reducing its speed-specific drag coefficient, $C_{Dfb} < C_{Dfa}$.

We consider the $n = 5$ parameters of this problem: the drag per unit width $F/w$ of the plate, it's length $l$, and bending rigidity $D$ as well as the uniform flow of velocity $U_\infty$ of a fluid with density $\rho$. Since this FSI problem can be described with $k = 3$ fundamental dimensions, namely length [m], time [s] and force [N] (keeping in mind that $1\text{N} = 1$ kg m/s$^2$), the Buckingham theorem (White, 2016 Chap. 5) tells us that we can formulate $n - k = 2$ dimensionless numbers to fully characterise this problem:

$$C_Y = \frac{C_D \rho l^3 U_\infty^2}{16D}, \quad R = \frac{F}{\frac{1}{2}\rho C_D w l U_\infty^2}, \qquad (6)$$

where $C_D$ is the drag coefficient of an equivalent rigid plate of the same width and length. These two dimensionless numbers are not unique, others could have been defined to characterise this problem. However, these two convey a physical interpretation. The Cauchy number $C_Y$ characterises the reconfiguration of an elastic structure under flow (de Langre, 2008; Gosselin *et al.*, 2010). It is proportional to the force the fluid flow would exert on the original undeformed shape of the structure over the bending rigidity of the structure. The reconfiguration number $R$ highlights the effect of flexibility on the drag. It is defined as the ratio of the drag on the flexible object over that of an equivalent rigid object. Another way to look at the reconfiguration number is to ask what length would a rigid equivalent plate need to have the same drag as the flexible one in the same conditions? Luhar and Nepf (2011) call this the effective length $l_e$. It is related to the reconfiguration number by $l_e/l = R$.

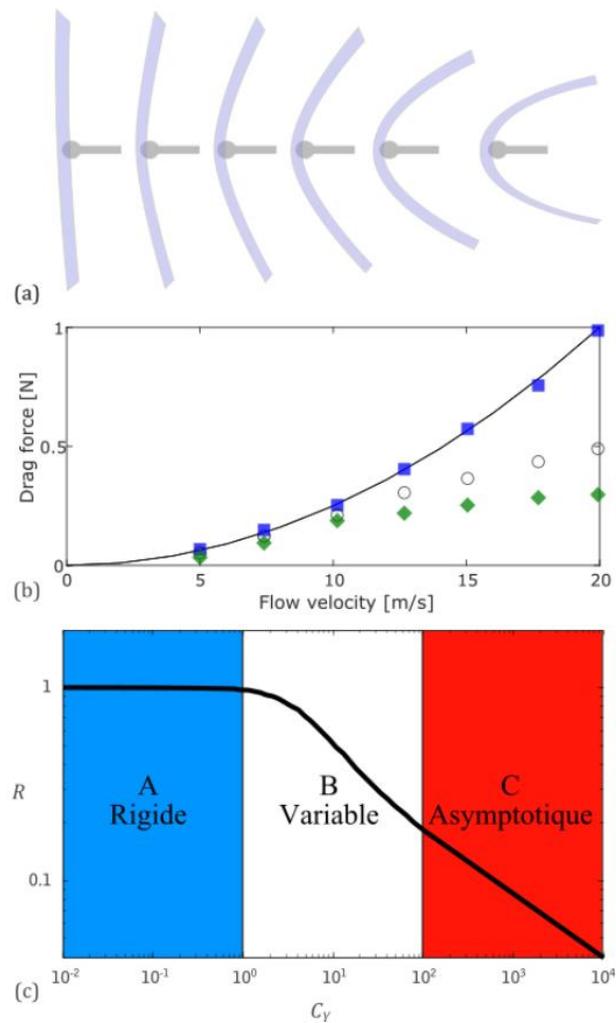

*Figure 2 Two-dimensional reconfiguration of a flat plate in a wind tunnel: (a) schematics showing snapshots of the reconfiguration for six velocities with increasing speed from left to right, the flow coming from the left bends the plate supported at its centre; (b) measurements of the drag on three plates of the same dimensions with very large rigidity (blue squares), small rigidity (white circles) and smallest rigidity (green diamonds); and (c) dimensionless representation of the drag of the flat plate modelled analytically highlighting the different regimes of reconfiguration. The blue region represents the small Cauchy regime, where the plate behaves as a rigid plate, the white region is a transition, and the red region represents the high Cauchy regime where the plate is highly deformed. Inspired by Gosselin et al. 2010.*

The proportionality relationship (4) can be rewritten in dimensionless form as $R \propto C_Y^{V/2}$. The great interest in using dimensional analysis is to fully characterise a problem and make it independent of units. It encompasses all notions of the plate being big or small, rigid or flexible, or the flow being fast flowing water or slow moving air. Gosselin *et al.* (2010) developed a semi-analytical model of the reconfiguration of the plate. The drag predictions of this model are shown in Figure 2 (c) in dimensionless form. There is only one master curve for the problem with three distinct regimes. All the data points of Figure 2 (b) would fall on this curve. At low $C_Y$ (A), the deformations are small and the drag on the plate varies like that of a rigid bluff body (i.e., $R \approx 1$ and $V = 0$). In this regime, the reconfiguration number is independent of the Cauchy number. For $C_Y \sim 1$ (B), reconfiguration becomes important and drag reduction occurs. In this regime, the Vogel exponent is variable. For $C_Y > 100$ (C), reconfiguration is so important that the original dimension of the plate becomes irrelevant in the scaling of the drag. This leads to a constant slope on the logarithmic plot of $R \propto C_Y^{-1/3}$ or equivalently, $V = -2/3$. This exponent can be explained in physical terms by performing an asymptotic analysis (Alben *et al.*, 2002; Gosselin *et al.*, 2010). In the asymptotic case of the wind deforming the plate so much that its initial length $l$ no longer characterises it, the two-dimensional reconfiguration of the plate with its drag per unit width $F/W$ becomes a problem of $n = 4$ parameters:

$$\frac{F}{w}[\text{N/m}], \quad D[\text{N·m}], \quad \rho[\text{kg/m}^3], \quad U_\infty [\text{m/s}].$$

With the same $k = 3$ fundamental dimensions as earlier, the Buckingham theorem tells us that we can reduce this problem into $n - k = 1$ single dimensionless number which cancels out all units

$$\frac{F}{wD^{1/3}\rho^{2/3}U_\infty^{4/3}}. \quad (7)$$

For dimensional similarity to hold, the proportionality between the parameters must remain. This can be enforced by equating this number to a constant. Thus, for a given plate and fluid, $w, D, \rho$ are constant and it holds true that $F \propto U_\infty^{4/3}$, or equivalently, $V = -2/3$. It is from the fact that the object is so deformed that its original length is no longer a part of the problem that this scaling arises. Note that this scaling is valid for the large deformation regime only (zone C in Figure 2 c). This point is important: the drag of a flat plate cannot be represented with a single Vogel exponent over the whole Cauchy number range. If we extrapolate this understanding to real plants, it means that they possibly experience different regimes of reconfiguration. It would thus not be appropriate to try to fit a single Vogel exponent for a whole data set of real plant drag spanning different regimes.

The dimensionless number representation also allows highlighting the difference and resemblances between different problems. The drag measurements on flat plates and filaments in a wind tunnel (Gosselin *et al.*, 2010; Gosselin and de Langre, 2011), and on flexible millimetre-long fibres in a soap-film flow (Alben *et al.*, 2002) can all be collapsed onto the same master curve which matches the model of Figure 2 (c). These experiments at different scales all exhibit the same form of reconfiguration with the same regimes and Vogel exponents. On the other hand, Schouveiler and Boudaoud (2006) presented experiments on a disk cut along a radius which rolls up into an increasingly acute cone as the water flow speed is increased as depicted in Figure 3, reminiscent of the tuliptree leaf of (Vogel, 1989). This tridimensional reconfiguration is geometrically and dimensionally different from that of the rectangular plate, thus it has a different drag scaling with a different Vogel exponent in its high deformation regime. A similar argument to that of the rectangular plate can be made for the disk. For the high Cauchy regime, the disk is rolled up in a cone so tight and so deformed that its original diameter $d$ does not influence its drag anymore. Thus, the problem of the drag on the reconfiguring disk is described with four parameters:

$$F[\text{N}], \quad D[\text{N·m}], \quad \rho[\text{kg/m}^3], \quad U_\infty\ [\text{m/s}],$$

which can be assembled in a single dimensionless number

$$\frac{F}{D^{2/3}\rho^{1/3}U_\infty^{2/3}}. \quad (8)$$

Similarly as for Eq. (7), for dimensional similarity to hold, the number above must be equal to a constant. Thus, $F \propto U_\infty^{2/3}$ or equivalently, $V = -4/3$. This large deformation regime scaling is equivalent to that found by Schouveiler and Boudaoud (2006) with their analytical model. Interestingly, Schouveiler and Eloy (2013) showed how an intact disk supported at its centre exhibits different modes of draping when subjected to normal water flow. In their experiments, drag is not measured, but since the bending occurs about radii of the initial disk, just like the rolling cone of Schouveiler and Boudaoud (2006), we could expect the same drag scaling. This has not been demonstrated. The difference between the dimensionless numbers of eqs (7) and (8) arises from the definition of a drag per unit width in the first case and simply drag in the second case. It is this difference in dimensions which leads to different scalings. Depending on how a leaf or another plant organ bends under the flow, we can expect its drag to scale differently.

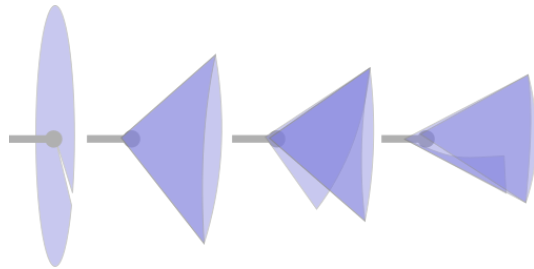

*Figure 3 Three-dimensional reconfiguration of a disk cut along a radius. The schematics depict the disk supported at its centre subjected to flow from left to right for four different flow speeds. With increasing flow speed, the disk rolls into a tighter cone. Inspired by Schouveiler and Boudaoud 2006.*

## Poroelasticity: effective velocity reduction

Whereas fluid must flow *around* solid objects like leaves and polymer sheets, it can flow *through* tree crowns, forests, or crop canopies, which are porous. For these porous and elastic—poroelastic—bodies, not only do the area reduction and the streamlining play a role in drag reduction, a third mechanism of effective velocity reduction can be important. We consider the tree of Figure 4 as a poroelastic structure from a macroscopic or whole-plant perspective. In (a), when subjected to a low-speed wind $U_{\infty a}$, the tree has a projected area $A_{fa}$, its drag-inducing elements (its leaves) have a drag coefficient $C_{Dfa}$ and the effective flow velocity they encounter is $U_{fa}$. At higher flow speed in (b), the cross-sectional area of the tree is reduced and its drag-inducing elements deform and become more streamlined ($C_{Dfb} < C_{Dfa}$). In addition to these two mechanisms, the flow profile through the tree crown is modified. As the tree crown deforms, its local porosity changes and the leaves face a modified effective velocity.

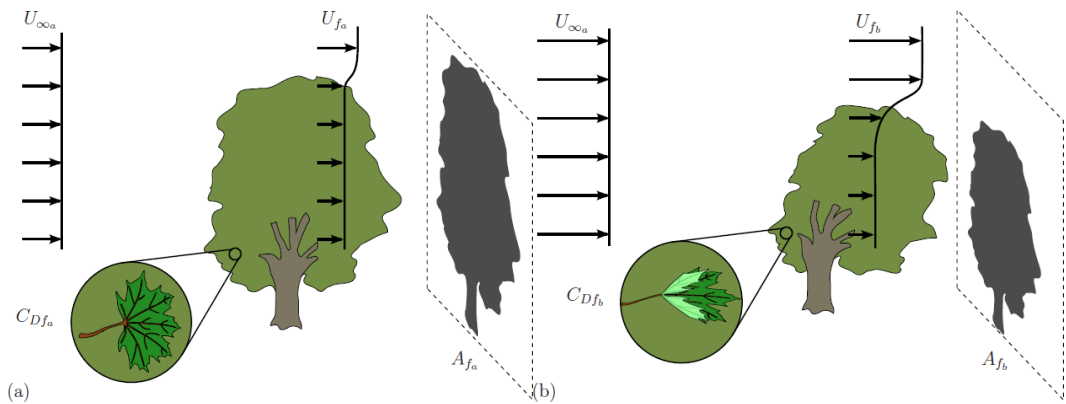

*Figure 4 Schematics of the drag reduction on a whole tree considered as a poroelastic structure subjected to a low-speed wind in (a) and a high-speed wind in (b).*

To reproduce the reconfiguration of the poroelastic tree of Figure 4 with an idealised system, Gosselin and de Langre (2011) measured the drag in a wind tunnel of a ball made of hundreds of rubber filaments attached together at its core. The reconfiguration of the poroelastic ball is schematised in Figure 5. The radially extending filaments bend downstream under the flow, aligning more and more with it as the velocity increases. Thus, the ball reduces its cross-sectional area, and becomes more compact. To characterise the reconfiguration of this ball, another dimensionless parameter is required in addition to its Cauchy number which can be formulated similarly to Eq. (6). We define the surface density as the ratio of the cross-sectional area of all the drag components of the porous body over the cross-sectional area of the undeformed poroelastic body, i.e.,

$$\gamma = \frac{N d_f \frac{d_b}{2}}{\frac{1}{4}\pi d_b^2} = \frac{2 N d_f}{\pi d_b}, \qquad (9)$$

where the poroelastic ball has a diameter $d_b$, the $N$ filaments composing the ball have a diameter $d_f$ and a length $d_b/2$. The surface density is a variation on the leaf-area index (LAI), which is the ratio of total projected leaf area per unit ground area. Here the area of drag-creating elements is projected on the cross-section of the poroelastic body normal to the flow. For small surface density values, we can expect the system to behave like an isolated structure. In this case, the structure is so porous, that the elements see an undisturbed flow. For large values of $\gamma$, drag-creating components become sheltered by one another inside the poroelastic structure. Gosselin and de Langre (2011) showed that dividing the Cauchy number by the surface density allows us to roughly coalesce the onset of reconfiguration in systems of different porosities. It amounts to dividing the aerodynamic load evenly on all the structural elements composing the porous bodies. Individual filaments or plates (both with $\gamma = 1$) and poroelastic balls start reconfiguring when the ratio $C_Y/\gamma$ reaches a value between 1 and 3. Both systems follow different $R = f(C_Y/\gamma)$ curves, but their onset of reconfiguration can be made to match by proper scaling. Gosselin and de Langre (2011) also presented a numerical model of poroelastic reconfiguration based on a conservation of momentum in the direction of the flow coupled with the large deformation Euler-Bernoulli equation of many beams. The model reproduces the drag measurements and reconfiguration of the poroelastic balls, and it allows interesting predictions in the asymptotic regime of large deformations ($C_Y \gg 1$). Whereas an isolated structure like a plate or a filament has a Vogel exponent of $V = -2/3$ in this regime, the Vogel exponent of a dense poroelastic structure ($\gamma > 10$) approaches $V \sim -1$, no matter the exact value of surface density. A dimensional argument can be made to explain this. For $C_Y \gg 1$, the poroelastic structure is so deformed that its original diameter $d_b$ is not important anymore. Moreover, the poroelastic body is so dense in drag-creating filaments, that the exact size of these filaments $d_f$ is not important for drag creation. The surface density is so large that the flow loses almost all its momentum in the porous body irrespective of the exact shape of the filaments. At high $C_Y$ and high $\gamma$, the problem of the drag of the poroelastic ball has four parameters

$$F[\text{N}], \quad EI[\text{N·}m^2], \quad \rho[\text{kg/m}^3], \quad U_\infty\ [\text{m/s}],$$

which can be assembled into a single dimensionless parameter

$$\frac{F}{(\rho EI)^{1/2} U_\infty}. \qquad (10)$$

For dimensional similarity to hold, $F \propto U_\infty$ or equivalently, $V = -1$. This scaling argument is interesting because it holds true under the assumption that the original dimensions of the poroelastic body and that of its drag-creating elements do not matter. Table 1 compares this Vogel exponent obtained for a synthetic system with those of best fits on wind tunnel drag measurements of coniferous trees (Mayhead, 1973; Vogel, 1989; Rudnicki *et al.*, 2004). The drag-creating elements of these trees are needles which bear a certain resemblance to the filaments of the synthetic poroelastic ball. Both the needles and the filaments behave structurally as beams. It might not be a coincidence that the Vogel exponents of these trees are similar to the synthetic system although this requires further investigation.

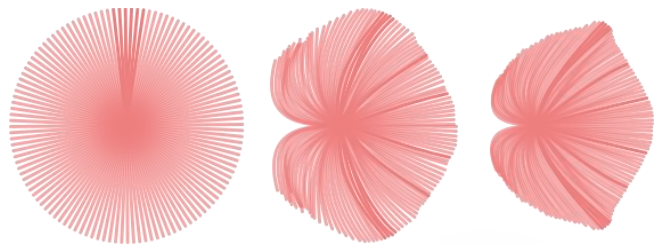

*Figure 5 Reconfiguration of a poroelastic sphere. The schematics depict the reconfiguration of a ball made of rubber hairs in a wind tunnel and subjected to three flow speeds. A no flow, the ball is round and the hairs point in all directions. With increasing flow speed, the hairs bend down and the ball becomes more dense. Inspired by Gosselin and de Langre 2011.*

The theoretical model of Gosselin and de Langre (2011) is based on a number of simplifications, notably in terms of flow simulation. To understand further and test the robustness of the $V = -1$ scaling for poroelastic structures, a more representative, higher fidelity model will need to be developed. Ricciardi *et al.* (2009) derived a fluid-structure interaction model to simulate the vibrations in nuclear reactor cores based on poroelastic modelling. The reactor core, made up of bundles of fuel rods with flowing water in between, is space-averaged to give rise to a homogenised continuum formulation. The local porosity and thus the local volume faction of water to solid fluctuates with the dynamics of the rod bundles. The complex geometry and space-varying orientation of the branches and leaves make this a much more complicated problem for plants than it is to simulate a regular array of cylinders in a nuclear reactor core, however the fundamental concept is the same. Eventually, an approach will have to be developed to space-average the tree architecture and its structure similarly to Lopez *et al.* (2015) and couple it with the homogenized fluid flow such as that of Zampogna *et al.* (2019).

In deriving these drag scalings for plates, disks and porous flexible structures, one's goal could be to offer better and simpler models for the interaction of vegetation with flowing fluid in larger simulations. For example, models exist to predict lodging and windthrow conditions on cereals and forests (Baker, 1995; Gardiner *et al.*, 2000; Berry *et al.*, 2003; Dupont *et al.*, 2010, 2015) and simulations are performed to reconstruct the conditions of destructive storms (Dupont and Brunet, 2006). However, although the results of the models and simulations are very sensitive to the aerodynamic loading values, the modelling of wind forces on the plants remains either approximate or empirical and based on experimental measurements made on a plant species tested in a range of speeds and specific conditions. However, as we see here, even if one could assume that the interactions between plants and flow occur in the large deformation regime (they do not always), slight differences in the mode of reconfiguration can change the drag scaling with flow velocity. It will not be possible to come up with a universal drag scaling for plants. On the other hand, the $V = -1$ found for poroelastic structures seems robust and might have a broad application range. It should be investigated further since how drag scales on flexible structures is at the basis of understanding plant-flow interactions.

*Table 1 Compilation of values of Vogel exponents for poroelastic structures and coniferous trees.*

| System | Condition | Vogel Exponent | Source |
|---|---|---|---|
| Synthetic poroelastic ball | $C_Y > 100$ | $-1$ | Gosselin and de Langre (2011) |
| *Pinus taeda*, 1 m high | $8 - 19 \, m/s$ | $-1.1$ | Vogel (1989) |
| *Pinus taeda*, branch | $8 - 19 \, m/s$ | $-1.1$ | Vogel (1989) |
| *Pinus sylvestris,* 5 m high | $8 - 19 \, m/s$ | $-0.72$ | Mayhead (1973) |
| *Pinus contorta,* 5 m high | $8 - 20 \, m/s$ | $\sim -1$ | Rudnicki et al. (2004) |
| *Tsuga heterophylla,* 5 m high | $8 - 20 \, m/s$ | $-1$ | Rudnicki et al. (2004) |

## Torsion and chirality

All the work presented above dealt exclusively with bending deformation. However, most plants deform with significant torsion as well as bending when subjected to flow (Niklas, 1992; Vogel, 1996). An elegant example of this is the stem of the daffodil, *Narcissus pseudonarcissus*, that twists with the wind to reorient its flower downwind and minimise its drag (Etnier and Vogel, 2000; Vogel, 2007). This mode of reconfiguration is favoured by the elliptical shape of the cross section and the anisotropy of the fibrous material of the stem. Many natural structures deform more easily in torsion than in bending because they have a very large ratio of bending to torsional rigidity:

$$\eta = \frac{EI}{GJ}, \quad (11)$$

where $EI$ is the bending rigidity combining the Youngs modulus and the second moment of inertia, and where $GJ$ is the torsional rigidity combining the shear modulus and the polar moment of inertia. Whereas a man-made circular cross-section metal bar has a ratio of $\eta = 1.3$ (assuming a Poisson ratio of 0.3), many leaf petioles have a ratio between 3 and 8, a tree trunk has a value around 7, and a value of 68 has been measured for the U-shaped banana leaf petiole (Ennos *et al.*, 2000; Vogel, 2007; Louf *et al.*, 2018). The fibrous construction of plants and the geometry of their cross-section make them much more prone to twist than bend (Faisal *et al.*, 2016). Hassani *et al.* (2016) devised an experimental framework to study the effect of twist on reconfiguration. They fabricated composite rods made of soft polyurethane foam reinforced with nylon fibres. The arrangement of the relatively stiff fibres in the round cross-section made it such that the rods were easier to twist than to bend, and also that they were easier to bend along one axis than the other. Put in the wind tunnel at different angles to the flow, such an initially straight rod bends and twists in three-dimensional space. Under increasing flow velocity, the rod bends and twist to align its easier bending direction with the flow. Hassani *et al.* (2016) presented an "equivalent Cauchy number" function of the twist-to-bend ratio $C_Y^*(\eta)$, which collapses the drag measurements on the composite rods placed at different angles in the flow. It highlights the role of torsion in reconfiguration which is that it facilitates reconfiguration by allowing the structure to reorient and to bend in its easier axis, but it does not change the drag scaling. Torsion has no influence of the Vogel exponent, it helps make the reconfiguration more two-dimensional.

Many plants exhibit an intrinsic twist, or chirality, e.g., at rest, cattail leaves can twist by a full turn or more from their root to their tip (Zhao *et al.*, 2015). Chirality increases the critical self-weight buckling length of an upright ribbon (Schulgasser and Witztum, 2004), i.e., for a certain length range, a flat ribbon buckles under its on weight, whereas a chiral one made of the same material and with the same cross-section can stay upright. Consider a flat upright ribbon with a rectangular cross-section. It bends more easily in one direction $EI_{min}$, than the other $EI_{max}$. We define the bending stiffness ratio

$$\chi = \frac{EI_{min}}{EI_{max}}. \qquad (12)$$

If the length of the ribbon is increased up to a critical value, gravity makes it buckle in the plane of $EI_{min}$. Now, if the ribbon is chiral and has intrinsic twist as depicted in Figure 6 (a), this critical length is increased because the ribbon must now buckle in a tridimensional helix. The increase in critical length can be evaluated analytically and is shown in Figure 6 (b). For practical matter, vertical leaves fit the flat ribbon limit ($\chi \ll 1$), and we can expect their critical length to be increased by ~25% for large intrinsic twist. Schulgasser and Witztum (2004) spliced leaves together to show that the safety factor against self-weight buckling of the leaves of cattail, *Typha domingensis*, and daffodil, *Narcissus tazetta* was minimal. If these leaves were not chiral, their weight would make them buckle.

Chirality allows upright flat leaves to grow taller, but how does it influence reconfiguration? Hassani *et al.* (2017) fabricated flexible polymer chiral ribbons by pre-twisting flat strips of plastic and placing them in the oven above the glass transition temperature of the material. Upon cooling, these ribbons adopted a stress-free chiral shape. Hassani *et al.* (2017) then proceeded to measure their drag as they reconfigured in a wind tunnel for different incidence angles. Whereas the untwisted ribbon exhibits much variation in drag depending on its relative orientation in the flow, intrinsic twist makes the reconfiguration of the ribbons more isotropic. A numerical model of chiral ribbon reconfiguration predicted that the root bending moment on the chiral ribbon depends on the angle at which the flow hits its. Considering the worst case (the flow direction which creates the largest bending moment), Hassani *et al.* (2017) compiled the variation of the root bending moment versus intrinsic angle for different ribbons with the simulated properties of eelgrass, *Zostera Marina,* and cattail leaf, *Typha angustifolia*. The plot of Figure 6 (c) shows that for both simulated ribbons, the maximum bending moment increases with intrinsic twist angle for small angles. The lightly chiral ribbon thus experiences a root bending moment more than twice as large as the flat one. However, for larger intrinsic twist, the bending moment comes back down. The best strategy for minimising drag-induced bending moment according to Figure 6 (c) is thus the non-chiral flat ribbon. However, as we saw in Figure 6 (b), the best strategy to grow taller and avoid buckling is to be chiral. The best compromise for a ribbon-like plant which seeks to grow tall and minimise its drag is thus to grow with large intrinsic curvature with more than a full rotation over its length ($> 360°$). This increases the buckling length (Figure 6 b), and decreases the root bending moment without reaching its absolute minimum (Figure 6 c).

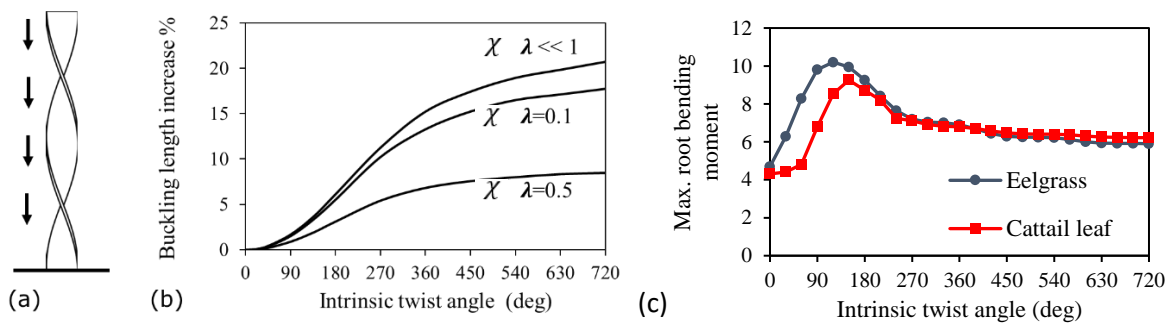

Figure 6 Buckling of a chiral ribbon under uniformly distributed self weight (a). Increase of the critical buckling length of a ribbon from intrinsic twist angle (b). The parameter $\chi$ is the ratio of bending rigidities in both planes of the ribbon cross-section (see Eq.

*(12). A value of $\chi \ll 1$ relates to a very flat and thin ribbon. Adapted from Schulgasser and Witztum (2004). Variation of the maximum dimensionless root bending moment computed for a chiral ribbon subjected to wind from all directions (c). The simulated ribbons had geometrical and material parameters values similar to those of eelgrass and cattail leaf. Data from Hassani et al. 2017.*

Some seaweeds also exhibit significant intrinsic twist. I have observed stipes of bull kelp, *Nereocystis luetkeana*, exhibiting many turns of twist. Their stipe resembling a licorice candy stick. It is not clear what purpose, if any, this chirality serves. The stipe is kept in tension by the pneumatocyst which acts as a float. Bull kelp runs no risk of buckling when submerged. Perhaps, this chirality is simply the accumulation of twist from the flow forcing: with time and growth, the stresses relax, and the twisted shape becomes permanent.

## Buoyancy

In the work discussed above, bending and torsional rigidity are the only restoring forces considered—the forces which oppose the deformation by the wind. In submerged vegetation, buoyancy can also act as a restoring force. To investigate the impact of buoyancy on reconfiguration, Luhar and Nepf (2011) constructed idealised model blades with two materials: silicone foam, and high density polyethylene (HDPE). Strips of different lengths of these materials were attached to the bottom of a flume one at a time, and their drag and reconfiguration was measured for increasing flow velocity. By varying the material and the strip length, they could vary the influence of buoyancy on reconfiguration. Their idealised system models a blade with uniformly distributed lacunae such as seagrass. Luhar and Nepf (2011) proposed the Buoyancy number to relate the restoring forces of buoyancy and that of bending stiffness

$$B = \frac{\Delta \rho g w h l^3}{EI}, \quad (13)$$

where the considered strip is of length $l$, thickness $h$, width $w$, bending rigidity $EI$, and density lighter than the surrounding fluid by $\Delta\rho$, and where $g$ is the gravitational acceleration. The first noticeable effect of buoyancy is that at a fixed value of Cauchy number, the drag reduction $R$ is less for the highly buoyant (large $B$) foam strip than for the neutrally buoyant HDPE one. Buoyancy resists the deformation by the flow and keeps the strip upright. It has the effect of delaying reconfiguration to higher $C_Y$ values. Notably, in the regime of large deformation ($C_Y > 2000$), the buoyancy number $B$ does not change the Vogel exponent. In this regime, the drag reduction $R(C_Y)$ curve becomes independent of $B$. Overall, buoyancy changes the values and slope of the drag reduction curve in the Variable regime (B) of Figure 2 (c) as well as the values of $C_Y$ delimiting this regime. However, the Rigid and Asymptotic regimes are left unchanged. Luhar and Nepf (2011) developed a semi-analytical model of the strip reconfiguring under the flow considering the buoyancy effect. The model reproduced both the reconfiguration of the synthetic specimens, and also the drag scaling of intertidal macroalga *Chondrus crispus* and terrestrial giant reed *Arundo donax* as measured by Boller (2006) and Harder *et al.* (2004), respectively.

## Skin friction

The reconfiguration discussed up to this point only considered form drag, also known as pressure drag. Friction drag can also be important on plants, notably on kelp which align with the incoming flow (Koehl and Alberte, 1988; Johnson and Koehl, 1994; Koehl *et al.*, 2008). Luhar and Nepf (2011) estimated roughly the validity of the assumption of neglecting skin friction drag. They reckon that it makes less than 10% of the total drag of a flexible reconfiguring blade if

$R > 6.8/\sqrt{Re}$ for laminar boundary layer,     (14)

$R > 0.05$ for turbulent boundary layer on smooth surface,     (15)

where $Re = U_\infty l/\nu$ is the Reynolds number of a flat plate aligned with the flow, and $\nu$ is the kinematic viscosity. Including skin friction in the semi-analytical model of blade reconfiguration similar to those of Gosselin *et al.* (2010) and Luhar and Nepf (2011), Bhati *et al.* (2018) showed that it imposes a minimum attainable reconfiguration number. Lowering the Reynolds number increases the skin friction and raises the minimum attainable reconfiguration number. However, it is not clear if those limits can be attained in a macroscopic system. Viscous effects being predominant on microfluidic systems at low Reynolds number, whereas flow inertia tends to dominate on plants.

## Dynamics of plants in flow

Wind and water current can make plants move and oscillate in many ways. In the following section, we look specifically at the loads which are generated on individual plants oscillating in flow.

### Wave action

Seaweeds growing in the intertidal zone, soft corals and kelp forests in the subtidal zone, and cordgrasses in salt marshes must all cope with wave action. Fluctuations of flow velocities introduce dynamics into the reconfiguration of plants. Based on the work of Blevins (1990), Gaylord and Denny (1997), Luhar and Nepf (2016), Leclercq and de Langre (2018), and Lei and Nepf (2019), we review the different mechanisms of interaction at play for plants in oscillatory flow, define proper dimensionless numbers to characterise this problem, present the different kinematics regimes and their associated drag scaling.

We neglect boundary layers and consider waves much larger in length than the structure they impact. Hence, the wavy flow can be modelled as a spatially uniform flow of time-varying velocity

$$U_\infty(t) = U_c + U_w \sin(\omega t), \quad (16)$$

where $U_c$ is the constant component of the flow velocity and $U_w$ is the oscillating component with frequency $\omega$ (in rad/s). To illustrate the fluid forces acting on a seagrass or an artificial blade, we first consider a solid (perfectly rigid) blade of length $l$, width $w$, and thickness $h$, where $l \gg w \gg h$. We consider the width of the blade at rest to be aligned with the direction of the oscillatory flow. The force per unit length exerted by the flow on the blade can be written as

$$\Delta F_{wc} = \rho w h \dot{U}_\infty + m_a \dot{U}_\infty + \frac{1}{2}\rho C_D w |U_\infty|(U_\infty), \quad (17)$$

where the raised dot $(\dot{\ })$ denotes a derivative with respect to time, and where $m_a$ is the added mass per unit length. The first term on the right-hand side of Eq. (17) is the virtual buoyancy force. The oscillation of the flow must be driven by a pressure gradient. This pressure gradient, just like pressure head caused by gravity, causes a force proportional to the displaced volume of fluid times the fluid density and the acceleration. This force has the same mathematical form as buoyancy—hence the name virtual buoyancy. The second term is the added mass—also referred to as virtual mass. When an object oscillates in flow (or equivalently when flow oscillates about a still object), it entrains a certain mass of fluid to oscillate with it. This added mass depends on the geometry of the object, its confinement, and the fluid density. The added mass of geometrically simple objects can be evaluated analytically. FSI textbooks list them for different geometries (Blevins, 1990; Naudascher and Rockwell, 2012). For example, a cylinder oscillating in still fluid perpendicularly to its axis has an added mass per unit length equal to the fluid density times the cross-sectional area of the cylinder. For the thin flat blade subjected to oscillatory flow aligned with its thickness, the added mass per unit length is approximately

$$m_a = \rho \pi w^2/4. \quad (18)$$

The third term on the right-hand side of Eq. (17) is the drag, which in this case is a fluctuating quantity because of the fluctuating flow velocity.

We introduce a new dimensionless quantity, the Keulegan-Carpenter number

$$K_C = \frac{2\pi U_w}{w\omega} = \frac{2\pi A_w}{w}, \quad (19)$$

where the wave amplitude $A_w$ corresponds to the excursion of a fluid particle over one wave. The Keulegan-Carpenter is a measure of "how transient" the flow over the object is and how important the drag is versus the added mass. We use this number to rewrite Eq. (17) in dimensionless form by dividing both sides of the equation by $\frac{1}{2}\rho C_D w U_w^2$

$$\frac{\Delta F_{wc}}{\frac{1}{2}\rho C_D w U_w^2} = \frac{4\pi}{C_D}\frac{h}{w}\frac{1}{K_C}cos(\omega t) + \frac{\pi^2}{C_D}\frac{1}{K_C}cos(\omega t) + \left|\frac{U_c}{U_w} + sin(\omega t)\right|\left(\frac{U_c}{U_w} + sin(\omega t)\right). \quad (20)$$

For a thin blade with $h \ll w$, which is a reasonable assumption for many aquatic plants, the virtual buoyancy term (first term on RHS of Eq. (20)) can be neglected. Leclercq and de Langre (2018) consider the reconfiguration of polymer blades in pure oscillatory flow with no steady current velocity component ($U_c = 0$). They find that the Keulegan-Carpenter number defines two regimes. For $K_C < 1$, the reconfiguration is flow-inertia dominated. The added mass term is more important than the drag one. For the Keulegan-Carpenter number to be small, the wave amplitude must be smaller than the width of the blade. In this regime, wave action can lead to resonance if their frequency ω is close to one of the natural frequencies of the blade. However, because the wave-amplitude must be small in this regime, they do not lead to large stresses. When $K_C > 1$, the reconfiguration is drag dominated. Flexibility allows drag reduction and the stresses due to fluid loading are concentrated at the root of the blade.

Since upright aquatic plants like seagrasses and reeds are much smaller in width than the excursion amplitude of the considered waves, the drag-dominated regime of $K_C > 1$ can typically be assumed. We consider this drag-dominated regime under combined current and wave action similarly to Lei and Nepf (2018). To account for this combined flow, we define three variations of the Cauchy number for current dominated, wave dominated, and mixed conditions

$$C_{Yc} = \frac{C_D \rho l^3 U_c^2}{16D}, \quad (21)$$

$$C_{Yw} = \frac{C_D \rho l^3 U_w^2}{16D}, \quad (22)$$

$$C_{Ywc} = \frac{C_D \rho l^3}{16D}\left(U_c + \frac{1}{2}U_w\right)^2. \quad (23)$$

The combined flow Cauchy number $C_{Ywc}$ is based on the time-averaged velocity of Eq. (16). To characterise the kinematics of narrow blade reconfiguration, Luhar and Nepf (2016) propose to compare the length of the blade with the oscillatory flow excursion length

$$L = \frac{l}{A_w}. \quad (24)$$

This blade to wave excursion length ratio and the velocity ratio $U_c/U_w$ allow defining the different kinematic regimes of drag-dominated reconfiguration. The parameter map of Figure 7 presents a synthesis of these regimes for the case where reconfiguration is important ($C_{Ywc} \gg 1$) based on the analysis of (Lei and Nepf, 2019). For $U_c/U_w > 2$, the current dominates and the reconfiguration is essentially static, recovering the classical scaling law of $R \propto C_{Yc}^{-1/3}$ found in Eq. (7). When $U_c/U_w < 0.25$, the flexible blade reconfiguration is wave-dominated. Its kinematic regime depends on the blade length with respect to the wave excursion amplitude. For $L > 1$, the waves are short, and the blade motion is small. The blade oscillates with small amplitude at the frequency of the waves. Even if the Cauchy number is large, the waves are too short in excursion to allow the blade to bend down parallel with the flow. In this case, by an argument of scaling the blade excursion with the wave excursion, Luhar and Nepf (2016) derive the following scaling $R \propto (C_{Yw}L)^{-1/4}$. In highly oscillatory flow $U_c/U_w < 0.25$ and long waves $L <$ 1, the blade is reconfiguring in a "quasi-static" way, following the alternating flow direction. In this kinematic regime, the drag on the flexible blade scales as $R \propto C_{Yw}^{-1/3}$. For the mixed flow conditions $0.25 < U_C/U_W < 2$ in Figure 7, the blade reconfigures in the direction of the current then oscillates about this equilibrium when $L > 1$, or reconfigures asymmetrically in a quasi-static way when $L < 1$. For all lengths $L$, Lei and Nepf (2018) show that a scaling of $R \propto C_{Ywc}^{-1/3}$ based on the combined wave-current Cauchy formulation is obeyed in the mixed flow conditions $0.25 < U_C/U_W < 2$.

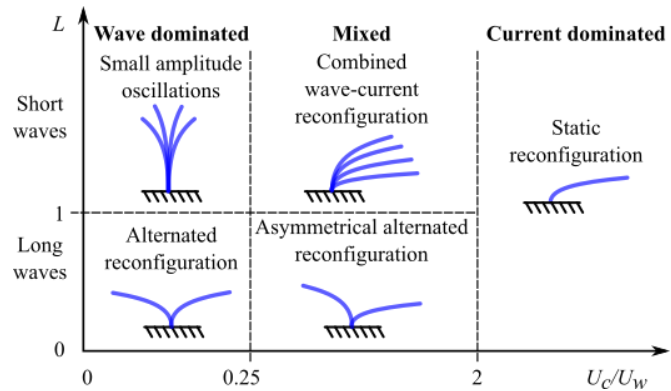

*Figure 7 Parameter map of the kinematic regimes of a flexible blade subjected to a flow with combined steady and oscillatory components. This map is valid for the regime of large wave-current Cauchy number. This map was created based on the results of Lei and Nepf 2018.*

The length parameter $L$ seems to allow differentiating the behaviours of real plants under flow, at least qualitatively. Koehl (1984) reports that sea palm algae, *Postelsia palmaeformis*, were completely stretched out by the successive peaks in acceleration and velocity of a wave hitting it. The longer algae Lessoniopsis litoralis on the other hand, were moving and aligning with the flow during the whole duration of the wave.

It is not clear how relevant the low Keulegan-Carpenter limit is for upright plants like weeds and reeds. In naval engineering, for ships and offshore structures, $K_C$ can still be small for large amplitude waves because the structures are so large. Thus, there can be large loads due to flow inertia. Small plants are drag-dominated by large amplitude waves. However, large kelps are subject to large inertia force and the effect of added mass on the stress they perceive is non-negligible (Utter and Denny, 1996; Gaylord and Denny, 1997; Denny *et al.*, 1997). There has been less work on idealised structures mimicking the reconfiguration of seaweed than that of terrestrial plants. Jensen and Denny (2015) tested and rejected the hypothesis that an "impingement" force plays a dominant role in the loading of seaweeds. They measured the drag of simple rigid objects suddenly hit by a fast water jet. They concluded that what is referred to as the impingement force is simply the sharp increases in drag, caused by brief spikes in water velocity at the instant the wave hits the seaweed.

From the start of this section, we assumed spatially uniform flows, because the wavelength of the incoming swell can typically be considered much larger than the scale of a plant. However, that is not always the case. Giant kelp and other seaweeds can grow tens of meters long. At a given instant, parts of a kelp blade can be pulled in opposite directions as it is longer than the wave. These opposite forces cancel each other out, and the stipe and holdfast do not have to withstand them. To account for the comparative size between the plant and the waves, FSI models like that of Denny *et al.* (1997) will need to be complexified to account for the finite size of the plant.

## Flutter

Reconfiguration and stability are intricately related. When exposed to wind, the tuliptree leaf, *Liriodendron tulipifera*, reconfigures by rolling into an increasingly acute cone which stabilises it. On the other hand, the leaves of a white oak, *Quercus alba*, flutter violently and tear at moderate wind speed (Vogel, 1989), even exhibiting a positive Vogel exponent—sign of a drag increase due to flexibility. With this perspective, it is interesting to see what understanding can be garnered from studying the stability of reconfiguring idealised structures.

In the static reconfiguration study of Gosselin *et al.* (2010), the wind tunnel experiments were limited in speed by a flutter instability. At high wind speed, the flat plate was reconfigured flat and was in large part parallel with the flow. Further increasing the wind speed would cause the plate to start flapping at high frequency like a flag. This would signal the end of the experiments as we sought static drag measurements. However, this raised an interesting question, which is the subject of the work of Leclercq *et al.* (2018): does flutter prevent drag reduction by elastic reconfiguration? They presented a reduced order numerical model of the dynamical effect of fluid flow on a beam deforming with large amplitude. Similarly to the work of Gosselin *et al.* (2010), their model predicted that a beam initially normal to the flow bends down with increasing Cauchy number, but instead of reconfiguring indefinitely, their model predicted a dynamic loss of stability as schematised in Figure 8 (a). They found that as they increased the Cauchy number, the beam transitioned from a static and stable reconfiguration regime, to a dynamic regime with periodic oscillations, and upon further increase of the Cauchy number, a non-periodic regime of oscillations would be reached. The exact transition between these different regimes depends on the mass ratio and the slenderness of the beam

$$\beta = \frac{m_a}{m+m_a}, \qquad \lambda = \left(\frac{2}{\pi}\right) C_D \frac{l}{w}, \quad (25)$$

where, $m$ is the mass per unit length of the blade, and where $m_a$ is its added mass, which can be evaluated with Eq. (18). The mass ratio tells us about how heavy or how light is the fluid vibrating with the blade. The mass ratio and the slenderness allow Leclercq *et al.* (2018) to formulate an answer about whether fluttering prevents drag reduction or not. These parameters influence the critical Cauchy number at which the beam starts fluttering and also how this flutter generates drag fluctuations. Increasing the values of $\beta$ or $\lambda$ increases the stability of the blade, i.e., it can reconfigure statically to higher values of the Cauchy number for higher values of $\beta$ and $\lambda$. For slender blades ($\lambda \geq 10$) and important added mass ($\beta > 0.5$), flexibility always leads to diminished perceived drag in comparison to an equivalent rigid beam. The instantaneous reconfiguration number of the beam fluctuates with its vibrations, but it never goes beyond 1, which would mean that the flexible structure perceives more drag than the rigid one because of fluttering. Moreover, for very slender blades ($\lambda \gg 1000$), the dynamical loads associated with fluttering are not that significant in comparison with the static drag. On the other hand, for a heavy beam with $\beta = 0.1$ and $\lambda = 10$, infrequent snapping events occurring on the fluttering beam cause spikes in the instantaneous drag higher than the static drag perceived by the rigid equivalent beam, i.e., peaks in the time-trace of the reconfiguration number beyond 2 were recorded. Flutter definitely limits how low the reconfiguration number can get, and for heavy beams, it can even lead to higher peak loads than on rigid beams.

The dynamics exhibited by the reconfiguring fluttering beam of Leclercq *et al.* (2018) is in many ways similar to that of the canonical fluid-structure interactions flag problem (Eloy *et al.*, 2007; Michelin *et al.*, 2008; Shelley and Zhang, 2011; Paidoussis, 2016). This problem has attracted a lot of interest from experimentalists and theoreticians, but most of the work is concerned with the kinematics. Authors describe the succession of oscillatory regimes encountered with increasing flow velocity, but few pay a lot of attention to the dynamical loads due to flutter. Yadykin *et al.* (2001) performed numerical simulations of the flutter of a long hanging ribbon in air. They showed that the maximum instantaneous drag on the ribbon increases gradually with the onset of flutter, but as the flow velocity is increased further and non-periodic motions are encountered, the drag can increase by more than an order of magnitude. These large loads probably correspond to the snapping events of Leclercq *et al.* (2018). However, since these simulations were for a heavy ribbon in air, it is not obvious that similar drag spikes would occur for a light kelp blade in water. Rominger and Nepf (2014) measured the dynamical drag on long strips of polymer behind a vortex-generating obstacle in a flume. These long strips mimicked the blades of *Macrocystis pyrifera*. The strip rigidity directly affected their dynamics. With decreasing rigidity, the strips oscillated more and perceived both larger time-averaged and instantaneous forces. Rominger and Nepf (2014) also developed an experimental method to measure how the flapping of the blades would affect their absorption of a chemical in the flume water. Increasing flexibility increases mass exchange.

## Vortices

The last topic we review here is the coupling of the dynamics of a plant organ with the vortices shed in its wake. Miller *et al.* (2012) investigated the reconfiguration of broad leaves in flood water. Leaves rolled into stable cones similarly to what Vogel (1989) observed in air flow. Upon imaging the flow behind the leaf using particle image velocimetry, Miller *et al.* (2012) saw a pair of stable contra-rotating vortices. To further understand the interplay between reconfiguration and stability, they simulated numerically the bidimensional flow behind a flexible beam perpendicular to the flow and mounted on flexible support. This flexible structure forming a T is illustrated in Figure 8. The structure deformed under the flow pressure, and eddies—or vortices—formed in its wake. The periodic shedding of vortices couples with the flexibility of the support and the mass of the structure to give rise to Vortex-Induced Vibrations (VIV). Miller *et al.* (2012) only observed these VIV on idealized T-shaped structures. The real leaves, despite their flexible petiole, did not vibrate as their conical reconfiguration promoted their stability.

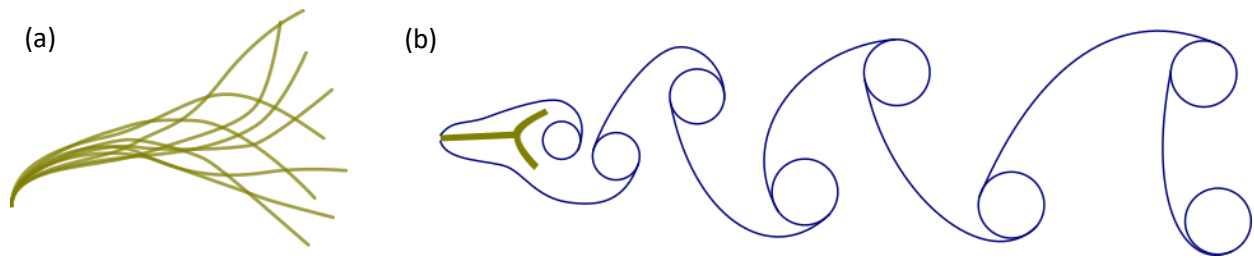

*Figure 8 Schematics of flow induced reconfiguring structures. A flat beam initially normal to the flow bends down and aligns with it before succumbing to a flag flapping instability (a). The different lines schematise instantaneous snapshots of the beam deformed shape, inspired by Leclercq et al. (2018). The flow around a flexible T-structure gives rise to a von Karman vortex street which induces vibrations (b), inspired by Miller et al. (2012).*

It is not clear if VIV can be responsible for significant loading on plants, but they can play a role to enhance particle interception. Oscillation of biological collectors can have significant effects on particle capture efficiency because it changes the momentum of the particles relative to the collector surface (Krick and Ackerman, 2015; McCombe and Ackerman, 2018). This was shown by field experiments on grass, as well as flume (flow chamber) experiments and computational fluid-dynamics (CFD) simulations. VIV are believed to cause these vibrations in grass stalks and helping them capture more pollen, thus favoring their reproductive success. Pollinating grasses must thus strike a delicate balance in flexibility to permit both vibrations and reconfiguration.

Vortex shedding occurs behind structures subjected to flow (Blevins, 1990; Paidoussis *et al.*, 2010). As vortices are shed from one side and then the other, surface pressure fluctuations are exerted on the structure. This oscillating pressure can cause vibrations of the structures, i.e., VIV. The Strouhal number is the dimensionless frequency of vortex shedding, i.e.,

$$S_T = \frac{fd}{U_\infty}, \qquad (26)$$

where $f$ is the frequency of vortex shedding in s$^{-1}$, and $d$ is the characteristic dimension of the object in the direction normal to the flow—for a cylinder, $d$ is the diameter. The Strouhal number defining the von Karman street behind a cylinder is dependent on the Reynolds number, but for a large range $10^2 < Re < 10^5$, the dimensionless frequency of vortex shedding is relatively constant around $S_T \sim 0.2$. This is a quick way to get a good indication whether vibrations are due to VIV or not. If one can measure the frequency of a flexible structure oscillating under flow, multiplying by its diameter and dividing by the flow velocity, the value of the Strouhal number should be around 0.2.

Soft corals face a physically similar problem to that of grasses collecting pollen from the air, they are suspension feeders and thus catch food particles brought in contact with them by the flow. On a SCUBA dive off Isla Mujeres near Cancun, Mexico, I was surprised by the vibrational behaviour of bipinnate sea plumes *Antillogorgia bipinnata*, which I did not find reported in the scientific literature. This soft coral grows to form colonies with an arborescent morphology. They are flexible and reconfigure passively in the ambient water currents. I observed *A. bipinnata* swaying back and forth at a low frequency under the forcing of wave action. However, the dynamics of the coral colony was richer than simple passive swaying: a high frequency vibration of the branches was clearly visible at a few hertz. No attention has been given in the literature to the nature nor the implications of this high frequency dynamics. As passive suspension feeders, *A. bipinnata* could potentially benefit from VIV by allowing the colony to sweep more water and reduce its boundary layer thickness, thus improving feeding rate.

The fact that corals, grasses and other plants benefit from VIV offers an exciting new paradigm in biomechanics and biomimetics. If VIV really offer a biological benefit, it makes one wonder if organism adapt their growth to favour them. Could the growth of an organism be tuned to ensure that it vibrates in its ambient flow? In engineering, VIV are a nuisance and the vast majority of research work on VIV thus pertains to minimising or eliminating them. As a matter of fact, the search query with the keywords "vortex vibrations suppression" results in 222 different inventions in Google Patents. They cause fatigue and can lead to failure of many engineering constructions such as bridges, chimney stacks, transmission lines, aircraft control surfaces, offshore structures, heat exchangers, and risers in petroleum production (Sarpkaya, 2004). Despite the scant literature on VIV in biomechanics, plants and corals show how VIV can be a feature rather than a nuisance. Taking inspiration from Nature's solution, engineers could develop system making use of VIV as a feature to develop energy harvesters, improve heat transfer of heat sinks and heat exchangers, design new filters, or improve any other device which relies on an efficient exchange of energy or mass with the ambient flow. Bernitsas *et al.* (2008) developed an energy harvesting device based on VIV, probably without the knowledge that plants have been benefiting from VIV as a feature for eons.

## Conclusion

Plants are highly flexible and thus significantly deform under water or air flows. This particularity made it necessary to branch out (pardon the pun) from the classical engineering FSI research and develop new analytical tools to consider these problems. The study of idealised ribbons, beams, and plates allows

Posted online 05/16/2019    Mechanics of a Plant in Fluid Flow    F.P. Gosselinunderstanding the fundamental mechanisms at play in plant reconfiguration and vibrations. However, the understanding these simplified structures bring is for the moment mostly qualitative. Cauchy numbers are hard to define for real trees (Whittaker *et al.*, 2013, 2015; Chen and Chen, 2017). What bending rigidity should be used to define the Cauchy number of a tree? That of its trunk, a branch, a petiole, a leaf? Ideas have been brought forward to use fractals to model the branching structure of a tree (McMahon and Kronauer, 1976; Rodriguez *et al.*, 2008; Eloy, 2011), but work is still necessary to link these fractals with real plant drag.

An obvious fertile ground for new research avenues in plant FSI is at the intersection between different phenomena. For example, Tadrist *et al.* (2018) determined the transition between leaf flutter and turbulence buffeting of branches. Other potential questions one might ask is how does static reconfiguration affect stability and how in return do vibrations affect the loads perceived by the plant? How can grasses or corals use vibrations to increase their pollen or food capture efficiency while still being flexible enough to reconfigure and cope with high winds or waves? How does elastic reconfiguration affects wind-induced pruning (Lopez *et al.*, 2011, 2014) or wind-induced pollen dehiscence (Timerman *et al.*, 2014)?

Plant canopies such as forests, crops and seagrass covers were purposely as their collective motions give rise to an entire new set of phenomena. Crops subjected to wind exhibit waves termed *honami* (Py *et al.*, 2006; Gosselin and de Langre, 2009), and seagrasses subjected to water flow exhibit similar waves termed *monami* (Ghisalberti and Nepf, 2002). Tree motions affect wind dynamics in storms (Dupont *et al.*, 2018). Emergent vegetation motions dampen incoming wave surges (Nové-Josserand *et al.*, 2018) and affect gaseous exchanges at the water surface (Foster-Martinez and Variano, 2016). All these collective phenomena have at their root, the interactions between flow and individual plants. Again, looking at the intersection between different phenomena looks like fertile ground for further research endeavours.

The work to be done is one of integration. Integrate different phenomena in a model and understand their interactions. Integrate these models developed on idealised structure with data measured on real plants. Bridge the two research approaches.

## Acknowledgements

We acknowledge the support of the Natural Sciences and Engineering Research Council of Canada (NSERC), [funding reference number 175791953].## Bibliography

**Alben S, Shelley M, Zhang J**. 2002. Drag reduction through self-similar bending of a flexible body. Nature **420**, 479–481.

**Alongi DM**. 2008. Mangrove forests: Resilience, protection from tsunamis, and responses to global climate change. Estuarine, Coastal and Shelf Science **76**, 1–13.

**Baker CJ**. 1995. The development of a theoretical model for the windthrow of plants. Journal of Theoretical Biology **175**, 355–372.

**Bernitsas MM, Raghavan K, Ben-Simon Y, Garcia EM**. 2008. VIVACE (Vortex Induced Vibration Aquatic Clean Energy): A New Concept in Generation of Clean and Renewable Energy From Fluid Flow. Journal of Offshore Mechanics and Arctic Engineering **130**, 041101–041101.Paper submitted to the Journal of Experimental Botany    21/26